\documentclass[aps,pra,reprint,superscriptaddress,showpacs]{revtex4-1}
\pdfoutput=1

\usepackage{amsmath}
\usepackage{graphicx}
\usepackage{epstopdf}
\usepackage{hyperref}
\hypersetup{
	pdfauthor={Gerard Higgins, Weibin Li, Fabian Pokorny, Chi Zhang, Florian Kress, Christine Maier, Johannes Haag, Quentin Bodart, Igor Lesanovsky, and Markus Hennrich},
	pdftitle={A single strontium Rydberg ion confined in a Paul trap},
	pdfsubject={Trapped Rydberg ion},
	pdfkeywords={Quantum Optics, Quantum Information, Rydberg states, Ion trapping, Intensities and shapes of atomic spectral lines, Stark effect in atoms}
}

\begin{document}

\title{A single strontium Rydberg ion confined in a Paul trap}

\author{Gerard Higgins}
\affiliation{Department of Physics, Stockholm University, 10691 Stockholm, Sweden}
\affiliation{Institut f\"ur Experimentalphysik, Universit\"at Innsbruck, 6020 Innsbruck, Austria}
\author{Weibin Li}
\affiliation{School of Physics and Astronomy, University of Nottingham, Nottingham, NG7 2RD, United Kingdom}
\affiliation{Centre for the Mathematics and Theoretical Physics of Quantum Non-equilibrium Systems, University of Nottingham, Nottingham, NG7 2RD, United Kingdom}
\author{Fabian Pokorny}
\affiliation{Department of Physics, Stockholm University, 10691 Stockholm, Sweden}
\author{Chi Zhang}
\affiliation{Department of Physics, Stockholm University, 10691 Stockholm, Sweden}
\author{Florian Kress}
\author{Christine Maier}
\author{Johannes Haag}
\affiliation{Institut f\"ur Experimentalphysik, Universit\"at Innsbruck, 6020 Innsbruck, Austria}
\author{Quentin Bodart}
\affiliation{Department of Physics, Stockholm University, 10691 Stockholm, Sweden}
\author{Igor Lesanovsky}
\affiliation{School of Physics and Astronomy, University of Nottingham, Nottingham, NG7 2RD, United Kingdom}
\affiliation{Centre for the Mathematics and Theoretical Physics of Quantum Non-equilibrium Systems, University of Nottingham, Nottingham, NG7 2RD, United Kingdom}
\author{Markus Hennrich}
\affiliation{Department of Physics, Stockholm University, 10691 Stockholm, Sweden}
\email[]{markus.hennrich@fysik.su.se}

%\date{\today}

\begin{abstract}
	Trapped Rydberg ions are a promising new system for quantum information processing. They have the potential to join the precise quantum operations of trapped ions and the strong, long-range interactions between Rydberg atoms. Technically, the ion trap will need to stay active while exciting the ions into the Rydberg state, else the strong Coulomb repulsion will quickly push the ions apart. Thus, a thorough understanding of the trap effects on Rydberg ions is essential for future applications. Here we report the observation of two fundamental trap effects. First, we investigate the interaction of the Rydberg electron with the quadrupolar electric trapping field. This effect leads to Floquet sidebands in the spectroscopy of Rydberg D-states whereas Rydberg S-states are unaffected due to their symmetry. Second, we report on the modified trapping potential in the Rydberg state compared to the ground state which results from the strong polarizability of the Rydberg ion. We observe the resultant energy shifts as a line broadening which can be suppressed by cooling the ion to the motional ground state in the directions orthogonal to the excitation laser.
\end{abstract}

\pacs{32.80.Ee, 37.10.Ty, 32.70.-n, 32.60.+i}
%{32.80.Ee = Rydberg states, 37.10.Ty = Ion trapping, 32.70.-n = Intensities and shapes of atomic spectral lines, 32.60.+i = Stark effect in atoms}
\maketitle

Trapped ions are one of the most mature implementations of a quantum computer. The trapped ion approach has set several benchmarks with qubit lifetimes up to minutes \cite{wineland1998}, entanglement operations with error probabilities smaller than $10^{-3}$ \cite{ballance2016,Gaebler2016}, and with up to 14 entangled qubits \cite{monz2011}. Trapped ions also assume a leading role in the implementation of quantum algorithms \cite{chiaverini2005a,brickman2005,Debnath2016,monz2016}, quantum error correction \cite{chiaverini2004,schindler2011,nigg2014}, and quantum simulations \cite{friedenauer2008,barreiro2011,lanyon2011,islam2013}.

The standard method to realize quantum information processing with trapped ions employs the common motion for entanglement operations between the ion qubits \cite{Cirac1995}. A current limitation of trapped ion quantum computation is the limited storage capacity as it becomes more difficult to perform entanglement operations in large ion crystals due to the increasingly complex motional mode structure. Possible schemes to reach larger quantum systems include segmented ion traps \cite{kielpinski2002}, ion-photon networks \cite{Monroe2014}, and, trapped Rydberg ions \cite{Mueller2008,Schmidt-Kaler2011}.

Trapped Rydberg ions are a novel quantum system. Here, the outermost electron of an ion is excited into Rydberg states far away from the atomic core. Due to large generated dipole moments, Rydberg ions are envisioned to sense each other by means of a dipolar interaction. The advantage of the Rydberg interaction is that it does not depend on the motional mode structure, thus it may be used in larger ion crystals for entanglement operations \cite{jaksch2000,Mueller2008,Li2013a}.  
A similar entanglement method has been demonstrated with neutral atoms \cite{wilk2010,isenhower2010,Labuhn2016}. In this sense trapped Rydberg ions promise to join the advantages of both technologies: they combine the strong dipolar interaction between Rydberg atoms with the precise quantum control and long storage times of trapped ions.

Recently, trapped Rydberg ions have been realized for the first time using a single-photon excitation of $\mathrm{^{40}Ca^+}$ ions with vacuum ultraviolet laser light at 122nm \cite{Feldker2015}. Also, selective manipulation of the ground state was combined with optical pumping via the Rydberg state \cite{Bachor2016}.

\begin{figure*}
	\includegraphics[width=0.95\textwidth]{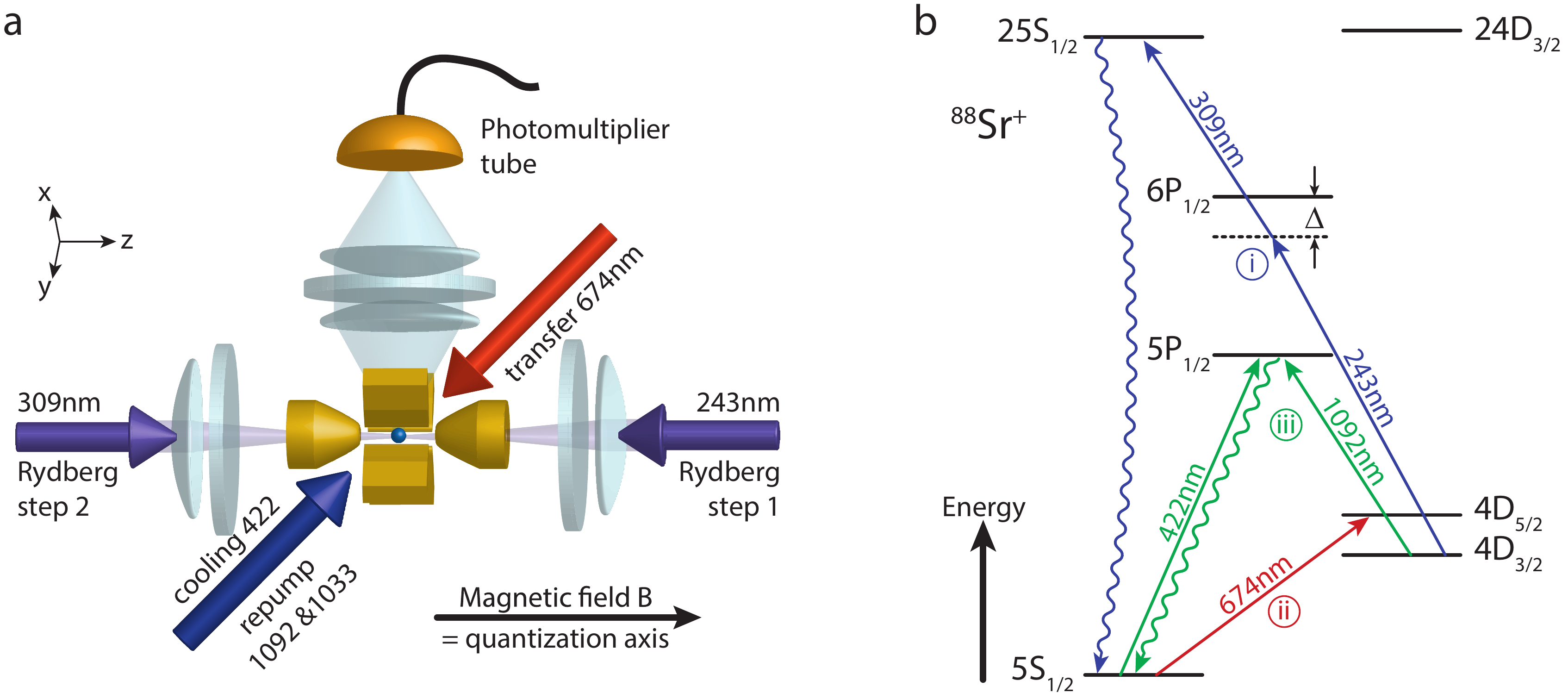}
	\caption{
		{\bf Experimental setup.} {\bf (a)} An ion is trapped in a linear Paul trap and manipulated by laser beams for Doppler cooling and fluorescence detection (422\,nm), repumping (1092\,nm and 1033\,nm), electron shelving (674\,nm), and Rydberg excitation (243\,nm and 309\,nm). A magnetic field with $\mathrm{B}=(0.3564 \pm 0.0008)\,\mathrm{mT}$ defines the quantization axis and is oriented parallel to the trap axis and the Rydberg excitation lasers. {\bf (b)} Energy level scheme of \textsuperscript{88}Sr\textsuperscript{+} and detection sequence of successful Rydberg excitation. Before Rydberg excitation the Doppler cooled ion is initialized in the metastable $\mathrm{4D_{3/2}}$ state via optical pumping. (i)~The Rydberg excitation lasers couple the initial $\mathrm{4D_{3/2}}$ state to a Rydberg S- or D-state, which then decays in multiple steps to the $\mathrm{5S_{1/2}}$ ground state with 95\% probability. (ii)~Any population in the ground state is transferred to the metastable $\mathrm{4D_{5/2}}$ state, allowing fluorescence detection to distinguish between successful Rydberg excitations (population in $\mathrm{4D_{5/2}\rightarrow}$ no fluorescence) and cases with no Rydberg excitations (population in $\mathrm{4D_{3/2}\rightarrow}$ fluorescence) in the final step~(iii). This sequence is typically repeated 100 times for each data point.\label{Fig1-setup}}
\end{figure*}
Here we report on a two-photon Rydberg excitation of \textsuperscript{88}Sr\textsuperscript{+} ions. The wavelengths used for Rydberg excitation (243\,nm and 305\,nm) are significantly easier to handle than the vacuum ultraviolet light used before. In particular, the laser photons contain significantly lower energy, thus no surface charging of the trap electrodes is observed. Also, in a two-photon excitation the lasers can be sent from opposite sides which puts the ions in an effective Lamb-Dicke regime and thus avoids Doppler broadening of the resonances. These advantages allow us to investigate the fundamental effects of the trap on Rydberg ions.

Modifications of the Rydberg properties due to the strong electric fields of the Paul trap have been predicted \cite{Mueller2008,Schmidt-Kaler2011}. One of the expected effects is that the Rydberg electron will interact with the quadrupolar electric trapping field. While Rydberg $\mathrm{S_{1/2}}$- and $\mathrm{P_{1/2}}$-states do not possess quadrupole moments due to their symmetry, and thus do not interact with the quadrupole field, higher angular momentum states will be affected. We experimentally investigate this fundamental effect, explore the different behaviour of $\mathrm{S_{1/2}}$- and $\mathrm{D_{3/2}}$-Rydberg states, and compare the experimental results to theoretical simulations.

A further novel property of Rydberg excited ions is that their trapping potential is modified compared to their ground state which can e.g.~induce structural phase transitions in an ion crystal \cite{Li2012}. This effect is caused by the strong polarizability of the Rydberg state which becomes polarized in the electric trapping field. Depending on the sign of the polarizability the induced dipole either weakens or increases the electric trapping field experienced by the ionic core and thus modifies the confining potential. We observe a first signature of this effect investigating a single Rydberg resonance. In particular, we see the modified trapping potential as a line broadening for a Doppler cooled ion with a thermal population distribution, as compared to a sideband cooled ion where most of the population resides in the motional ground state.

\section{Experimental system\label{ExpSystem}}
A single \textsuperscript{88}Sr\textsuperscript{+} ion is confined in a linear Paul trap and excited to Rydberg S- and D-states by a two-photon process as shown in Fig.~\ref{Fig1-setup}.
243\,nm laser light drives the first step from the metastable state $\mathrm{4D_{3/2}}$ to $\mathrm{6P_{1/2}}$, while tunable laser light at 304-309~nm excites the second step to reach $\mathrm{S_{1/2}}$- and $\mathrm{D_{3/2}}$-Rydberg states.

During Rydberg excitation the ion is confined in the electric field of a linear Paul trap. Note that the Rydberg ions do not get ionized by the electric trapping fields, as ions are generally held at zero electric field in the trap center. Being charged particles they will move to equilibrium positions where the electric fields of the trap and the neighbouring ions compensate. The remaining trapping potential forms an electric quadrupole field with the ion at the centre, at least given electric stray fields leading to micromotion of the ion are properly compensated. For further details on the experimental setup and the micromotion compensation see Appendices \ref{appendix:RydbergLasers} and \ref{appendix:IonTrap}.

\section{Ionic Rydberg states in a linear Paul trap\label{Theory}}
In the following we provide the theoretical background concerning the Rydberg excitation of a strontium ion held in a linear Paul trap. This will allow us to undertake a comparison between experimentally obtained and theoretically calculated excitation spectra. Specifically, it will enable us to identify and quantify the influence of the trapping field on the Rydberg-ion level structure.

The electric potential of the Paul trap reads
\begin{equation}
\Phi({\bf r},t)=\alpha\cos\Omega t\,(x^2-y^2)-\beta\,(x^2+y^2-2z^2)
\label{epaul}
\end{equation}
where $\alpha$ and $\beta$ are electric field gradients and $\Omega$ is the frequency of the radio-frequency (RF) electric field. In contrast to low-lying states, the weakly-bound Rydberg electron can exhibit a sizable coupling to the trap electric field. To illustrate this, we write the coupling Hamiltonian $H_{\text{et}}(\mathbf{r},t)$ as~\cite{Schmidt-Kaler2011}, 
\begin{eqnarray}
\label{etrap}
H_{\text{et}}(\mathbf{r},t)&=& {er^2}\left[-2\sqrt{\frac{\pi}{5}}\beta\,Y_2^0(\mathbf{\theta,\phi}) \right.\nonumber \\
& & \left.-\sqrt{\frac{8\pi}{15}}\alpha \cos\Omega t\,Y_2^{2}(\theta,\phi)+\text{H.c.}\right],\\
\nonumber
\end{eqnarray}
where $\theta$ and $\phi$ are polar and azimuthal angles with respect to the trap axis $z$, and $Y_{l}^m(\theta,\phi)$ are spherical harmonics. 

\begin{figure}
	\includegraphics[width=\columnwidth]{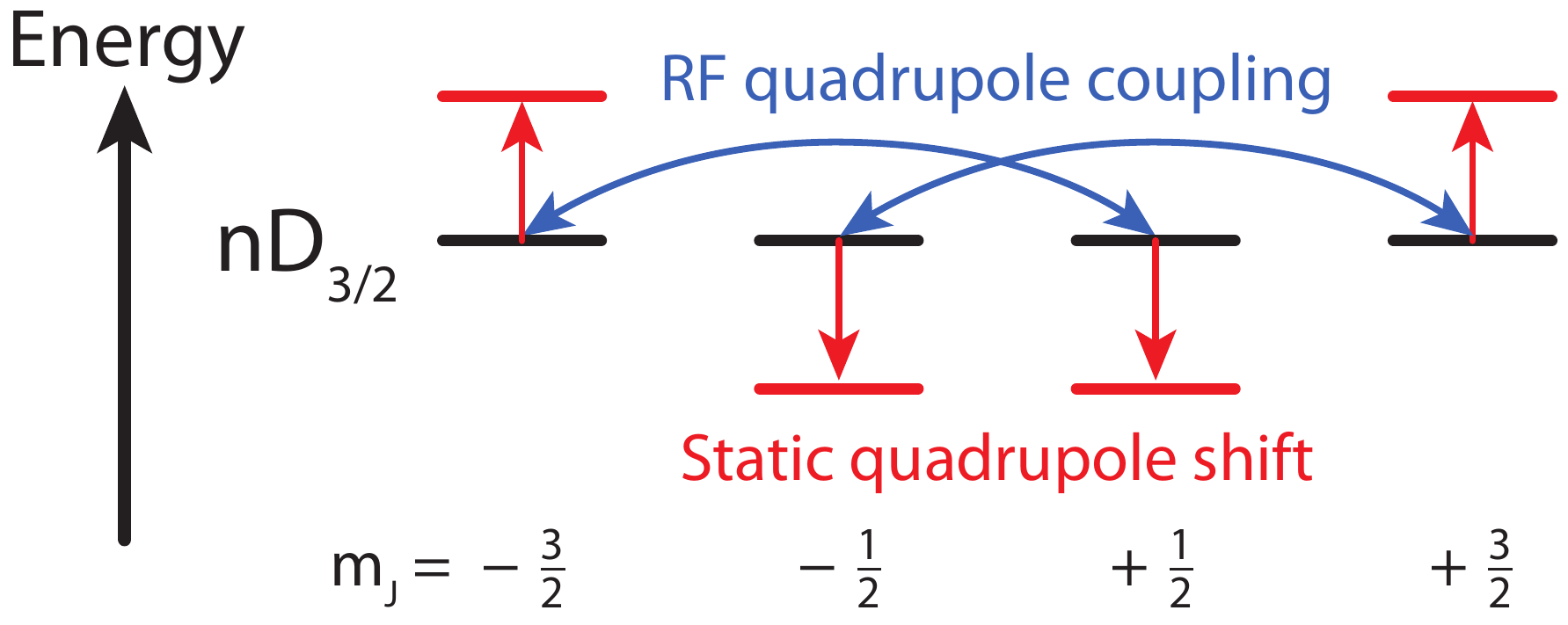}
	\caption{{\bf Electron-trap interaction.} Quadrupolar energy shifts and radio-frequency (RF) coupling due to the electron-trap interaction $H_\mathrm{et}$ for an electron in Rydberg $nD_{3/2}$ state. 
		\label{ET-interaction}}
\end{figure}

For a Rydberg state $|nLJm_J\rangle$, with $n,\, L,\, J$ the principal, angular, and total angular quantum numbers, and $m_J$ the projection of $J$ on the quantization axis (along the trap $z$-axis), the quadrupole coupling is non-zero when $J>1/2$, while it vanishes when $J=1/2$ due to selection rules. This shows that there is no first order effect for Rydberg $S$-states ($L=0$ and $J=1/2$). However, the coupling becomes significant in Rydberg $nD_{J}$ states ($J=3/2$ or $J=5/2$), as illustrated in Fig.~\ref{ET-interaction}. 
Specifically, for $J=3/2$ and magnetic field $B \| z$ the static part of the trap potential causes an energy shift to state $|nD\frac{3}{2}m_J\rangle$,
\begin{eqnarray}
\label{eq:trapstatic}
E_{\text{s}}&=&\frac{2}{5}(-1)^{|m_J|-1/2}\beta Q_{3/2}.
\end{eqnarray}
where $Q_J= -e\langle nDJ|r^2|nDJ\rangle$ denotes the corresponding quadrupole transition moment. For $n\gg 1$, its approximate value is  $Q_J\approx  \frac{e a_0^2 n^2}{2(2Z_e+1)}[5n^2+1-3L(L+1)]$, with core charge $Z_e=2$ and Bohr radius $a_0$. From Eq.~(\ref{eq:trapstatic}) it follows that the energy shifts for $|m_J|=1/2$ and $|m_J|=3/2$ have the same strength but opposite signs, as depicted in Fig.~\ref{ET-interaction}. 

The RF dependent part of Hamiltonian (\ref{ET-interaction}), on the other hand, couples different Zeeman states with $\Delta m_J=\pm 2$, see Fig.~\ref{ET-interaction}. As the RF frequency is much smaller than the fine structure splitting ($\Omega\sim$ MHz), we restrict the coupling to within the same Zeeman manifold (for states with identical quantum numbers $n$, $L$ and $J$). The RF dependent part of Eq.~(\ref{ET-interaction}) then assumes the form
\begin{eqnarray}
H_{\text{RF}}={\hbar C\cos\Omega t} \sum_{m_J=1/2}^{3/2}\biggl[|nLJ(m_J-2)\rangle\langle nLJm_J| + \text{H.c.}\biggr],\nonumber
\end{eqnarray}
where the constant $C=-2 Q_J\alpha/5\sqrt{3}\hbar$ is the effective Rabi frequency of the RF field.
	
In the current experiment, typical trap parameters are $\alpha\approx  3\times10^8\,\mathrm{V m^{-2}}$, $\beta\approx  6\times10^5\,\mathrm{V m^{-2}}$ and $\Omega\approx 2\pi\,18$ MHz. This yields a static frequency shift $|E_{\text{s}}|/\hbar\approx 2\pi\,43$\,kHz and effective Rabi frequency $C \approx 2\pi\,12$\,MHz in the Rydberg state $|\mathrm{24D}\frac{3}{2}m_J\rangle$. Note that the latter is comparable with the RF frequency, $C\sim\Omega$. An emerging feature is that Floquet sidebands will be populated~\cite{friedrich_theoretical_2005}, as the rotating wave approximation is not applicable in the quadrupole coupling Hamiltonian $H_{\text{RF}}$.

All subsequent calculations of the Rydberg spectra (shown in Figs.~\ref{Exp_25S} and \ref{ryd_ex_24D}) are performed with coupling Hamiltonian of the form (\ref{ET-interaction}). We take into account the experimental trap geometry and laser parameters (see Appendix~\ref{appendix:laser-ion} for the laser-ion interaction). The laser excitation dynamics is described through a quantum master equation.
 
\section{Electron-trap interaction acting on D- vs S-Rydberg states\label{ZeemanStructure}}
\begin{figure*}
	\includegraphics[width=0.85\textwidth]{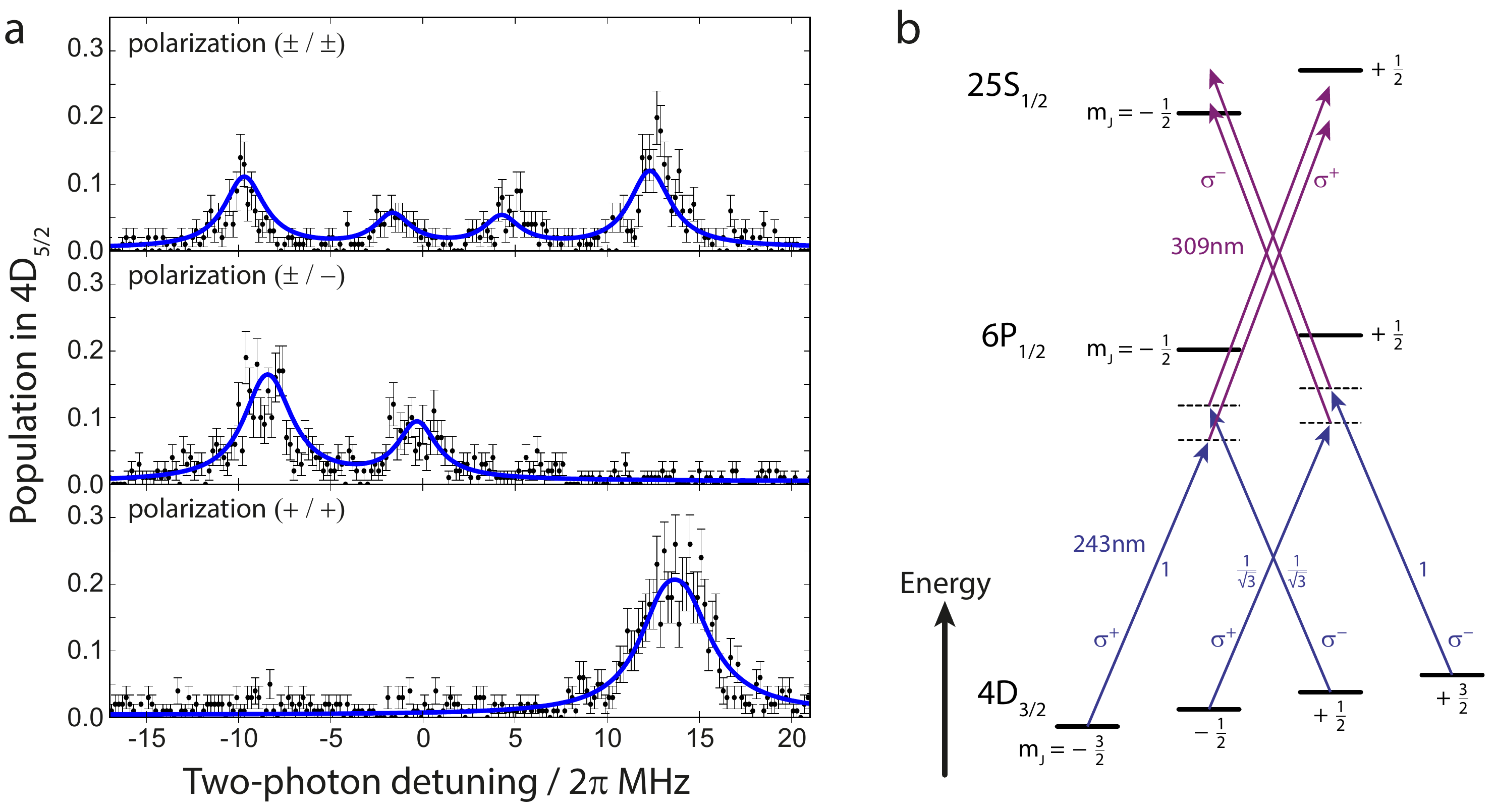}
	\caption{
		{\bf Zeeman splitting in S-Rydberg states.} {\bf (a)} Two-photon excitation spectra to state $\mathrm{25S_{1/2}}$. Successful Rydberg excitation is signalled by a high probability for shelving the ion to state $\mathrm{4D_{5/2}}$.
		The excitation spectrum shows four resonance peaks for both lasers in an equal superposition of $\sigma^+$ and  $\sigma^-$ polarization (denoted as $(\pm / \pm)$), two peaks when the second step 309\,nm laser is changed to $\sigma^-$ $(\pm / -)$, and a single peak when both lasers are $\sigma^+$ $(+ / +)$. Black dots are the measured data points with error bars due to quantum projection noise, blue lines are the simulated Rydberg excitation spectra (Rabi frequencies $\Omega_{243}=2\pi\,0.47$\,MHz, $\Omega_{309}=2\pi\,49$\,MHz, dephasing of the Rydberg state $\delta\omega_{25S}=2\pi\,2.2$\,MHz). 
		Off-resonant scattering from the intermediate state and spontaneous decay from the initial $\mathrm{4D_{3/2}}$ state each contribute to the background signal.
		{\bf (b)} Allowed transitions for Rydberg excitation to $\mathrm{25S_{1/2}}$. The Rydberg-excitation beams are aligned with the direction of the applied magnetic field at the position of the ion, thus electric dipole transitions which preserve the magnetic quantum number ($\mathrm{\pi}$-transitions) are not excited. With this constraint, only four non-degenerate transitions between the $\mathrm{4D_{3/2}}$ and $\mathrm{25S_{1/2}}$ Zeeman sublevels remain. As the frequency of the 309nm laser is scanned, each of the four transitions comes into resonance at a different frequency. 
		\label{Exp_25S}}
\end{figure*}

According to the previous considerations Rydberg S-states are expected to not interact with the electric trapping field, thus we expect an excitation spectrum with a simple structure. The experimental results for $\mathrm{25S_{1/2}}$ are shown in Fig.~\ref{Exp_25S}.
Peaks in the $\mathrm{4D_{5/2}}$ population result from the excitation of the ion to $\mathrm{25S_{1/2}}$ Rydberg state. Depending on the laser polarizations used, one, two or four peaks are observed. The observed resonance lines can be easily explained by the Zeeman splitting in the applied magnetic field of $B=(0.3564 \pm 0.0008)\,\mathrm{mT}$ where each of the four equally populated initial Zeeman levels couples to exactly one Rydberg level, see Fig.~\ref{Exp_25S}b. The simulation results corroborate this explanation. Relative amplitudes agree with the difference in Rabi frequencies due to the respective Clebsch-Gordan coefficients. Thus, the spectroscopy of the Rydberg S-state can be fully explained by the Zeeman effect in the applied magnetic field. No effect of the electron-trapping field interaction is observed.

\begin{figure*}
	\includegraphics[width=\textwidth]{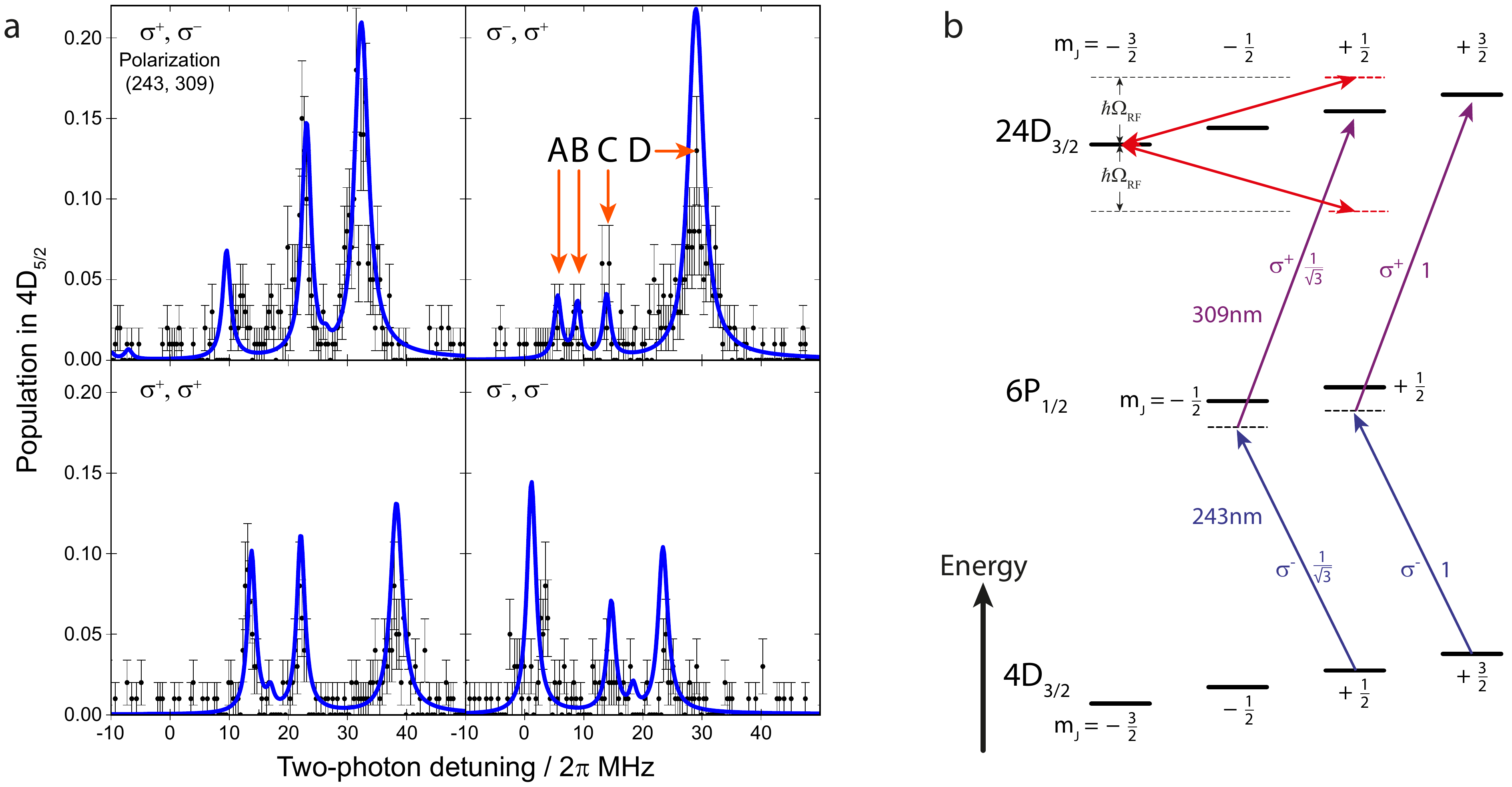}
	\caption{
		{\bf Electron-trap interaction for D-Rydberg states.} {\bf (a)} Excitation spectra and simulation results for $\mathrm{24D_{3/2}}$ (trapping parameters 
		$\{\omega_{\text{axial}},\omega_{\text{radial 1}},\omega_{\text{radial 2}}\}=2\pi\{(254\pm3),(600\pm10),(760\pm10)\}$\,kHz, radio-frequency drive $\Omega=2\pi\,18.153$\,MHz). The experimental data (error bars due to quantum projection noise) match the simulation results (dark blue line) when the electron-trap interaction is taken into account (simulation parameters Rabi frequencies $\Omega_{243}=2\pi\,0.09 $\,MHz, $\Omega_{309}=2\pi\,135$\,MHz, detuning $\Delta_{243}=+2\pi\,160$\,MHz, dephasing of the Rydberg state $\delta\omega_{\mathrm{24D}}=2\pi\,4.1$\,MHz). {\bf (b)} Energy level scheme explaining the RF sidebands in the excitation spectra. Sidebands appear at $\pm$ the RF trapping frequency $\Omega$ with respect to a neighbouring Zeeman state $\Delta m_J = \pm 2$. The resonances are additionally shifted due to the AC Stark effect by the Rydberg excitation laser at 309\,nm. \label{ryd_ex_24D}}
\end{figure*}

The effect of the electron-trap interaction becomes evident in the excitation spectra of the $\mathrm{24D_{3/2}}$ Rydberg state, see Fig.~\ref{ryd_ex_24D}. 
We observe a multitude of resonances with Floquet sidebands caused by the RF coupling between Zeeman sublevels within the $\mathrm{24D_{3/2}}$ manifold. 
The simulation matches the experimental data when the electron-trap interaction [Eq.~(\ref{ET-interaction})] is included.
Also, matching the positions of the AC-Stark shifted resonances in the simulation and the experimental data allows the Rabi frequency of the 309\,nm laser to be determined to within 10\,\%.

The observed resonance lines can be explained in simple terms as depicted in Fig.~\ref{ryd_ex_24D}b. Due to the RF coupling of the electron-trap interaction, sidebands at $\pm \Omega$ relative to neighbouring Zeeman levels ($\Delta m_J=\pm 2$) become visible. For high Rabi frequencies the lines are offset by AC Stark shifts. The resonance frequencies can be identified by a diagonalization of the coupling Hamiltonian within the $\mathrm{24D_{3/2}}$ manifold, see Appendix \ref{appendix:24DResonancesTheory}. For instance for $\sigma^-$/$\sigma^+$ polarizations for the first/second Rydberg excitation steps and the Rabi frequencies as in the matching simulation, we obtain four eigenfrequencies at $2\pi\,\{5.8, 9.2, 14.1, 29.0\}$\,MHz detuning with corresponding eigenstates
\begin{eqnarray}
|A\rangle&=&+0.72~|\frac{1}{2}, 0\rangle - 0.69~|\frac{3}{2}, +1\rangle.\nonumber \\
|B\rangle&=&-0.21~|\frac{3}{2}, 0\rangle + 0.98~|\frac{1}{2}, +1\rangle \nonumber \\
|C\rangle&=&+0.69~|\frac{1}{2}, 0\rangle + 0.72~|\frac{3}{2}, +1\rangle \nonumber \\
|D\rangle&=&+0.97~|\frac{3}{2}, 0\rangle + 0.22~|\frac{1}{2}, +1\rangle\nonumber
\end{eqnarray}
where we use $|m_J,n\rangle=|\mathrm{24D} \frac{3}{2} m_J,n\rangle$, with $n$ being the number of quadrupole excitations of the Floquet sidebands. The corresponding states $|A\rangle$ to $|D\rangle$ are marked in the top right panel of Fig.~\ref{ryd_ex_24D}. In these states, the "absorbing" states $|m_J, -1\rangle$ have negligible contributions, since the electron-trap interaction is further detuned, see Fig.\ref{ryd_ex_24D}b.

\section{Modified trapping potential in the Rydberg state\label{SingleResonance}}
The strong polarizability of the Rydberg state is expected to modify the effective trapping potential of the ion. The change in the radial trapping potential of the Rydberg state compared to lower-lying states is \cite{Li2012,Mueller2008,Schmidt-Kaler2011}
\begin{equation}
V_{\text{add}} = - \left(\alpha^2 + 2 \beta^2\right) \mathcal{P}_{n,L}\,\rho^2    \approx    - \alpha^2 \mathcal{P}_{n,L}\,\rho^2,
\end{equation}
where $\rho$ is the radial center-of-mass coordinate of the ion, and $\mathcal{P}_{n,L}$ is the polarizability of the Rydberg state with quantum numbers $n$, and $L$. 
The polarizability scales as $\sim n^7$, thus the influence of $V_{\text{add}}$ should increase for higher and higher Rydberg states.

\begin{figure*}
\includegraphics[width=0.8\textwidth]{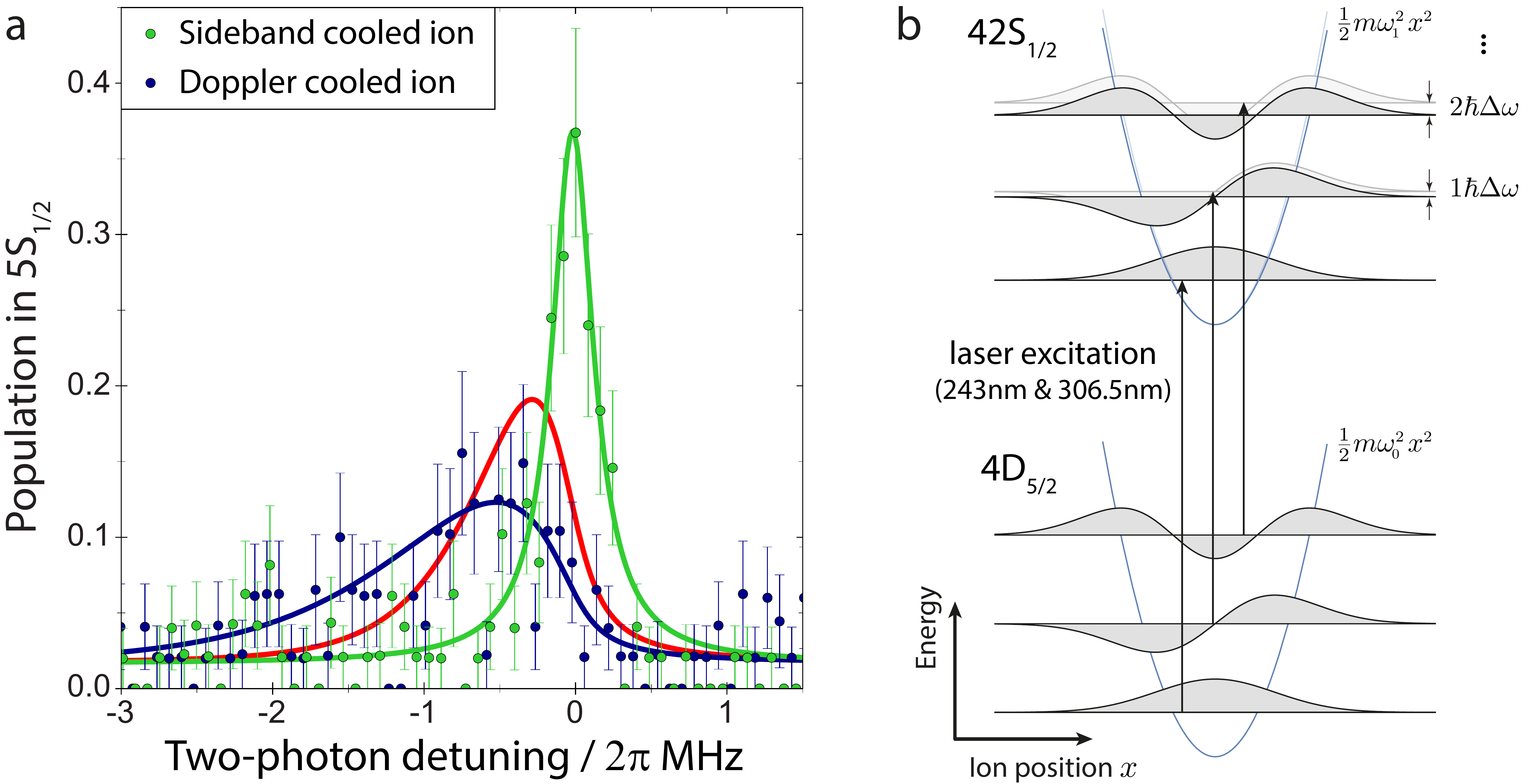}
\caption{
	{\bf Modified Rydberg trapping potential.}
	{\bf (a)} Observation of a single Rydberg resonance starting from the state $\mathrm{4D_{5/2}}$ $m_J=-\frac{5}{2}$ to the Rydberg state $\mathrm{42S_{1/2}}$ $m_J=-\frac{1}{2}$ (trapping parameters $\{\omega_{\text{axial}},\omega_{\text{radial 1}},\omega_{\text{radial 2}}\}=2\pi\{(872\pm5),(1660\pm30),(1720\pm30)\}$\,kHz, radio-frequency drive $\Omega=2\pi\,18.153$\,MHz). With sideband cooling we observe a single resonance with linewidth $2\pi\,(300\pm50)\,$kHz mainly limited by the laser linewidths. With only Doppler cooling the resonance is shifted lower in energy, has smaller amplitude and asymmetric shape with $\approx 2\pi\,(1.4\pm0.1)\,$MHz linewidth. The model curves represent a single resonance with $2\pi\,(300\pm50)\,$kHz linewidth for the sideband cooled case, and a thermally broadened line with $\Delta \omega=-20$\,kHz (red) [$\Delta \omega=-42$\,kHz (blue)] for the Doppler cooled case. 
	{\bf (b)} Energy scheme of the motional state during Rydberg excitation. Due to the different trapping potential in the Rydberg state compared to the ground state the laser excitation is shifted out of resonance for higher motional quantum numbers.\label{ryd_ex_from_4D5h_SB_cooling}}
\end{figure*}

We now investigate the effect of the modified trapping potential by analyzing the Rydberg excitation spectrum of state $\mathrm{42S_{1/2}}$. The controlled preparation of the ion in a single Zeeman sublevel combined with driving only a single transition in the Rydberg excitation leads to the observation of a single Rydberg resonance as depicted in Fig.~\ref{ryd_ex_from_4D5h_SB_cooling}a. The details on state preparation and detection of Rydberg excitation in this context are given in Appendix \ref{appendix:SingleRes}. 
After Doppler cooling we observe a linewidth of $2\pi\,(1.4\pm 0.1)$\,MHz, which is significantly broader than the expected natural linewidth calculated from quantum defect theory of $2\pi\,39$\,kHz. However, with radial sideband cooling the observed linewidth is reduced to $300$\,kHz. This sideband cooled linewidth can be fully explained by natural and laser linewidths ($\approx 2\pi\,200\,$kHz).

Commonly, one would suspect the broadening to be caused by the Doppler effect, however, we do not observe a reduced linewidth after axial sideband cooling (along the direction the excitation laser propagates). However, we observe a reduced linewidth after radial sideband cooling (perpendicular to the laser propagation direction). This effect can be explained by the modified trapping potential in the radial directions. 
For the $\mathrm{42S_{1/2}}$ Rydberg state quantum defect theory predicts a polarizability $\mathcal{P}_\mathrm{42S}=17.6\times 10^9$\,atomic units. As a result the trapping frequency in radial direction in the Rydberg state, $\omega_1$, should be smaller than the trapping frequency in a low-lying state, $\omega_0$, by $\Delta\omega=\omega_1-\omega_0\approx -20$\,kHz.
Consequently, the motional energy levels are closer in the Rydberg state than in low-lying states, and the vibrational wavefunctions are modified, see Fig.~\ref{ryd_ex_from_4D5h_SB_cooling}b. 
When the Rydberg excitation drives transitions between the low-lying state and the Rydberg state, the transition frequency is shifted depending on the initial motional state. In particular, the resonance frequency for phonon number conserving transitions is shifted by $n \Delta\omega$, when the ion has $n$ radial phonons, see Fig.~\ref{ryd_ex_from_4D5h_SB_cooling}b. Since the wavefunction in Rydberg and low-lying state are not perfectly orthogonal, the phonon number is not necessarily conserved during Rydberg excitation. Nevertheless, as the trapping potentials in our case are still very similar, so are the vibrational components of the wavefunctions. Thus, the Franck-Condon factors for the Rydberg excitation are approximately given by Kronecker-delta functions $\delta_{n,m}$ and population should be mainly transferred to quantum states with the same motional quantum number. In principle, with sufficiently high resolution one should be able to identify individual lines separated by $\Delta \omega$ with relative amplitudes due to the thermal population. In our experiment we cannot resolve the splitting for $\mathrm{42S_{1/2}}$ since our lasers are broader than the shifts. Nevertheless, we observe a red-shifted, asymmetrically broadened line due to thermal population of the two radial modes of motion, see Fig.~\ref{ryd_ex_from_4D5h_SB_cooling}a. 

We model the asymmetric lineshape in Fig.~\ref{ryd_ex_from_4D5h_SB_cooling}a taking into account the measured thermal population of the two radial modes after Doppler cooling of $\langle n_x\rangle = 13.4 \pm 0.5$ and $\langle n_y\rangle = 8.9 \pm 0.4$. The measured data in Fig.~\ref{ryd_ex_from_4D5h_SB_cooling}a seems to deviate from this model for the theoretical value of $\Delta\omega=-20$\,kHz. If we use $\Delta\omega$ as a free fitting parameter, we reach good agreement for $\Delta\omega=-42$\,kHz.

The polarizability of the Rydberg state $\mathrm{42S_{1/2}}$ critically depends on its energy difference to the neighbouring Rydberg P-states, which up to now can only be predicted theoretically due to the lack of experimental data. Thus, a value of $|\Delta \omega|$ larger than expected, could mean that the P-Rydberg states directly above ($\mathrm{42P_{1/2}}$ and $\mathrm{42P_{3/2}}$) might be closer to $\mathrm{42S_{1/2}}$ than predicted. 

\section{Conclusion\label{Conclusion}}
We have investigated several elementary trap effects of Rydberg ions, which will be essential for future applications in quantum technologies. In particular, we have observed the effect of the electron-trap interaction on Rydberg states with different symmetries. While Rydberg S-states are unaffected by the electron-trap interaction, Rydberg D-states interact with the quadrupolar trapping field. In the future the quadrupole moment of Rydberg D-states might lead to new types of interactions like quadrupole-charge, quadrupole-dipole or quadrupole-quadrupole interactions that could be used for directional coupling in quantum information tasks. Nevertheless, Rydberg S-states remain the less intricate system and might thus be easier to control.

Moreover, we have been able to deterministically control the initial state for Rydberg excitation which has allowed us to investigate a single Rydberg resonance. Electron-ion coupling has become visible as an increase in the Rydberg resonance linewidth of a Doppler cooled ion compared to a sideband cooled ion. As predicted, the large polarizability of the Rydberg states in the electric field of the Paul trap leads to different trapping potentials in low-lying and Rydberg-excited states, and to broadening of the resonances for the Doppler cooled case. The resultant weaker or stronger localisation of a Rydberg ion might be used in the future for quantum information processing to cut longer ion chains into smaller sub-units for localised quantum operations \cite{Li2013} or for implementing exotic interactions through vibrational mode shaping \cite{Nath2015}. 
A sideband cooled ion instead is largely unaffected by the electron-ion coupling due to its precise localisation. Thus, sideband cooling might be required for coherent Rydberg excitation and to avoid unwanted entanglement between the electronic state and the ion motion during Rydberg excitation. Microwave-dressed Rydberg states could be used to equalise the trapping potentials as proposed for realising efficient Rydberg quantum gates \cite{Li2013a}. Such Rydberg gates would make trapped Rydberg ions a novel system for quantum information processing.

\appendix

\section{Rydberg excitation lasers\label{appendix:RydbergLasers}}
The 243\,nm laser light for the first Rydberg-excitation step is produced in a commercial system, in which 970\,nm infrared laser light from a diode-laser pumped tapered-amplifier system is frequency-quadrupled to 243\,nm. Similar systems are used for the two-photon excitation of hydrogen \cite{Kolachevsky2006}.
Tunable 304-309\,nm laser light for the second Rydberg-excitation step is produced in two stages. In the first stage two infrared photons, from a 1551\,nm diode-laser fibre-amplifier system and a tunable 998-1030~nm diode-laser pumped tapered-amplifier system, are combined by sum-frequency generation in a periodically-poled lithium niobate crystal \cite{Wilson2011}.
The resultant 608-618\,nm laser light is frequency-doubled to produce tunable 304-309\,nm laser light, which covers the wavelength range for excitation of Rydberg states from principal quantum number n=24 up to the second ionization threshold.
The first Rydberg-excitation laser is frequency stabilized to $\mathrm{\approx}$100\,kHz linewidth (in-loop estimate) by locking the 970\,nm fundamental to a reference cavity. The second Rydberg-excitation laser is frequency stabilized also to $\mathrm{\approx}$100\,kHz linewidth (in-loop estimate) by referencing 608-618\,nm laser light to a cavity and applying feedback to the 998-1030\,nm fundamental laser.

The lasers are sent from opposite sides along the trap axis. The counter-propagating beams significantly reduce thermal effects in the two-photon Rydberg excitation. The effective Lamb-Dicke parameter for the two-photon transition is $\eta=0.044$ at an axial trapping frequency of $\omega_z=2\pi\,872$\,kHz. Thus, after Doppler cooling the ion resides within the Lamb-Dicke regime and Doppler broadening can be neglected.

The Rydberg-excitation lasers are sent through hydrogen-loaded, single-mode photonic crystal fibres. Such fibres offer laser mode cleanup and stable beam pointing while resisting ultraviolet solarization \cite{Colombe2014}. The lasers are focused by two commercial achromat lenses which image the $10\,\mathrm{\mu m}$ diameter fibre core with unity magnification onto the ion. The laser beams are sent through holes in the end-cap electrodes and up to 120\,nW (16.8\,mW) of 243\,nm (309\,nm) laser light is focused to $(5.1\pm0.8)\,\mathrm{\mu}$m ($(6.8\pm 1.7)\,\mathrm{\mu}$m horizontally, $(4.9\pm 1.0)\,\mathrm{\mu}$m vertically) beam waist at the position of the ion.

\section{Ion trap\label{appendix:IonTrap}}
The ion is confined in a macroscopic linear Paul trap with titanium electrodes in a sapphire mount. The trap consists of four blade electrodes for radial confinement and two end-cap electrodes with optical access holes for axial confinement. 
Since Rydberg atoms may be extremely sensitive to electric fields (polarizability $\mathcal{P}_n~\sim~n^{7}$) with $n$ the principal quantum number of the Rydberg state, the ions are confined very close to the electric quadrupole null to minimize any detrimental effects of the field to the stability of Rydberg ions. Using the ``cross-correlation" and the ``resolved sideband" techniques \cite{Berkeland1998,Keller2015} micromotion is minimized and the residual electric field at the position of the ion is estimated to be $(3^{+12}_{-3})$\,Vm\textsuperscript{-1}.
The trap electrodes are electroplated in gold (work function $\mathrm{\approx}$5.3\,eV) to avoid the emission of photoelectrons if ultraviolet laser light hits the electrode surfaces (243\,nm photons carry 5.1\,eV energy).
Photoelectron emission causes time-varying stray electric fields and thus worsens the residual electric field at the ion position. Due to the gold-coating, the micromotion compensation parameters do not change over weeks of working with the ultraviolet Rydberg-excitation lasers.

\section{State preparation and detection of a single Rydberg resonance\label{appendix:SingleRes}}
A single isolated Rydberg resonance is accessible in the Rydberg excitation from the initial state $\mathrm{4D_{5/2}}$ to the Rydberg state $\mathrm{42S_{1/2}}$. This excitation scheme is interesting from a quantum information perspective, since $\mathrm{4D_{5/2}}$ may be used together with the $\mathrm{5S_{1/2}}$ ground state to store an optical qubit. The Rydberg excitation could be used for entanglement operations between two such optical qubits.

For state preparation, a Doppler (or sideband) cooled ion is initiated in the $\mathrm{4D_{5/2}}$ $m_J=-\frac{5}{2}$ Zeeman sublevel. First any population in $\mathrm{4D_{5/2}}$ is returned to the ground state using the 1033\,nm ``repump" laser and the ion is Doppler cooled. The 674\,nm ``qubit" laser, which drives the $\mathrm{5S_{1/2}\leftrightarrow 4D_{5/2}}$ transition, has a narrow linewidth (\textless600~Hz) which allows transitions between specific Zeeman sublevels to be individually addressed. By alternately driving the $\mathrm{5S_{1/2}~m_{J}=+ 1/2 \rightarrow 4D_{5/2}~m_{J}=- 3/2}$ transition and removing population from $\mathrm{4D_{5/2}}$ using the 1033\,nm ``repump" laser, population may be optically pumped to the $\mathrm{5S_{1/2}~m_{J}=- 1/2}$ Zeeman sublevel. Next population is transferred from the initial $\mathrm{5S_{1/2}}$ Zeeman sublevel to a specific $\mathrm{4D_{5/2}}$ Zeeman sublevel using the narrow 674\,nm ``qubit" laser. The fluorescence detection lasers are then turned on to check whether the population transfer was successful.

The detection of successful Rydberg excitation from initial state $\mathrm{4D_{5/2}}$ is simpler than for $\mathrm{4D_{3/2}}$, as the initial state $\mathrm{4D_{5/2}}$ can be directly distinguished from the final state $\mathrm{5S_{1/2}}$ by fluorescence detection without any need for additional shelving pulses. The Rydberg excitation and detection sequence is as follows. (i) Both Rydberg lasers are turned on, coupling $\mathrm{4D_{5/2}}\,m_J=-\frac{5}{2}$ via the intermediate state $\mathrm{6P_{3/2}}\,m_J=-\frac{3}{2}$ to $\mathrm{42S_{1/2}}\,m_J=-\frac{1}{2}$. According to quantum defect theory calculations 95\% of the population in $\mathrm{42S_{1/2}}$ quickly decays to the $\mathrm{5S_{1/2}}$ ground state. (ii) Finally fluorescence detection is used to distinguish between successful Rydberg excitations (population in $\mathrm{5S_{1/2}\rightarrow}$ fluorescence) and cases with no Rydberg excitation (population in $\mathrm{4D_{5/2}\rightarrow}$ no fluorescence).

\section{Laser-ion interaction Hamiltonian\label{appendix:laser-ion}}
The Hamiltonian used to simulate the laser-induced two-photon transition to Rydberg S- and D-states including the magnetic field-induced Zeeman effect is~\cite{Mueller2008,Schmidt-Kaler2011,Li2013a},
\begin{eqnarray}
\label{Hcen}
H &=& H_{\text{e}}+H_{\text{et}}(\mathbf{r},t)+H_{\text{B}}+H_{\text{L}}\\
\label{he}
H_{\text{e}} &=& \sum_{\mathbf{L}}\varepsilon_{\mathbf{L}}|\mathbf{L}\rangle\langle \mathbf{L}|, \\
\label{hie}
H_{\text{B}}&=& -\frac{e}{2m_{\text{e}}c}|B_z|(L_z+2S_z), \\
\label{hlaser}
H_{\text{L}} &=& \sum_{j}-e\mathbf{r} \cdot \hat{\epsilon}_j E_{j}\cos(\mathbf{k}_j\cdot\mathbf{R}-\omega_jt),
\end{eqnarray}
where $H_{\text{e}}$, $H_{\text{et}}(\mathbf{r},t)=-e\Phi({\bf r},t)$, $H_{\text{B}}$ and $H_{\text{L}}$ stand for the Hamiltonian for the valence electron, electron-trapping field coupling, Zeeman effect and laser-electron interaction. In the Hamiltonian, $\mathbf{r}$ and $m_{\text{e}}$ are the position and mass of the electron, $\mathbf{R}$ the centre-of-mass position of the ion, and $\mathbf{L}$ the multi-index quantum number $\mathbf{L}=\{n,\,L,\,J,\,m_J\}$. $\varepsilon_{\mathbf{L}}$ is the energy in the electronic state $|\mathbf{L}\rangle$. $e$ and $c$ are the elementary charge and speed of light in vacuum. $B_z$ is a static magnetic field parallel to the $z$-axis. $S_z$ and $L_z$ are the $z$-components of the spin and angular momentum operators. $E_j$ and $\hat{\epsilon}_j$ are the electric field and polarization of the $j$-th laser, whose wave vector and frequency are $\mathbf{k}_j$ and $\omega_j$. 

In the experiment, the first laser couples low-lying states $|\mathrm{4D}\frac{3}{2}m_J\rangle$ (denoted by $|g_{m_J}\rangle$) and $|\mathrm{6P}\frac{1}{2}m_J\rangle$ (denoted by $|e_{m_J}\rangle$) and the second laser couples the state $|\mathrm{6P}\frac{1}{2}m_J\rangle$ and Rydberg state  $|R_{m_J}\rangle$.  We neglect the sideband transitions, as the two lasers are counter-propagating and the effective Lamb-Dicke parameter $\eta<0.1$. Using the relevant electronic states as bases, the laser-ion interaction can be expressed as
\begin{eqnarray}
H_{\text{L}}&=&\sum_{\mathbf{M}}\left[\hbar\Omega_{1,\mathbf{M}}\cos\omega_1t|e_{m_J}\rangle\langle g_{m_J'}|\right. \nonumber\\
&+&\left.\hbar\Omega_{2,\mathbf{M}}\cos\omega_2t|R_{m_J}\rangle\langle e_{m_J'}| +\text{H.c}\right],
\end{eqnarray}
where  $\mathbf{M}=\{m_J,m_J'\}$ is a two-index number, and Rabi frequencies $\Omega_{1,\mathbf{M}} = -eE_1\langle e_{m_J}|\mathbf{r}\cdot \hat{\epsilon}_1|g_{m_J'}\rangle/\hbar$ and $\Omega_{2,\mathbf{M}} = -eE_2\langle R_{m_J}|\mathbf{r}\cdot \hat{\epsilon}_2|e_{m_J'}\rangle/\hbar$ depend on respective electronic states and laser polarization. 

It shall be pointed out that both micromotion and the trap field mediated Rydberg electron-ion coupling are relatively weak~\cite{Mueller2008,Schmidt-Kaler2011,Li2012}. Both effects are not experimentally resolved for low lying Rydberg states $n<30$, as polarizability and temperature of the ion are small, and micromotion is carefully compensated. Therefore we do not consider these effects in the numerical simulation. However the electron-ion coupling is observed for the higher lying Rydberg state $\mathrm{42S_{1/2}}$ as discussed in section \ref{SingleResonance}.
 
\section{Spectra of the Rydberg $\mathrm{24D_{3/2}}$ state\label{appendix:24DResonancesTheory}}
Here we provide a simple theory to explain the spectra shown in Fig.~\ref{ryd_ex_24D}a. First, we note that electronic states will be completely specified once laser polarizations are given. This allows us to omit the labelling of the quantum number $m_J$ in the electronic low-lying states. The two Rydberg states that are coupled by the quadrupole field will be labelled by $|R_1\rangle$ and $|R_2\rangle$ for convenience. Upon applying rotating-wave approximations to the laser induced transitions, the Hamiltonian to describe the Rydberg excitation dynamics [see Eq.~(\ref{Hcen})] becomes
\begin{eqnarray}
H_{\text{D}}&=& \hbar\Delta_{e}|e\rangle\langle e| +\sum_j \hbar\Delta_j|R_j\rangle\langle R_j| +H_{\text{RF}}\\ &+&\frac{\hbar}{2}\left[\Omega_{\text{l}}|e\rangle\langle g|+\Omega_{\text{u}}|R_1\rangle\langle e| +\text{H.c}\right], \nonumber
\end{eqnarray}
where we have assumed that the polarization is chosen such that state $|e\rangle$ couples to Rydberg state $|R_1\rangle$. $\Delta_e=\left(E_e-E_g\right)/\hbar - \omega_1$ and $\Delta_j=(E_j-E_g)/\hbar -\omega_1-\omega_2$ ($j=1,2$) give detuning of the electronic transition with respect to the laser frequencies, where the energy $E_s=\varepsilon_s + E^{(B)}_s$ ($s=g,e,1,2$) takes into account of both the electronic energy $\varepsilon_s$ and Zeeman shift $E^{(B)}_s$.

As $|\Delta_e|$ is typically larger than other quantities in the Hamiltonian, we can adiabatically eliminate state $|e\rangle$, which yields
\begin{eqnarray}
H_{\text{D}}&\approx& -\frac{\hbar\Omega_{\text{l}}^2}{4\Delta_e}|g\rangle\langle g| +\hbar\left( \Delta_1  - \frac{\Omega_{\text{u}}^2}{4\Delta_e}\right)|R_1\rangle\langle R_1|  \nonumber \\ 
&+&\hbar\Delta_2|R_2\rangle\langle R_2|+\frac{\hbar}{2}\left[\Omega'|R_1\rangle\langle g| +\text{H.c.}\right]  \nonumber \\ 
&+&H_{\text{RF}},
\label{eq:Hdsim1}
\end{eqnarray}
where $\Omega' = -\Omega_{\text{l}}\Omega_{\text{u}}/2\Delta_e$ is the two-photon  Rabi frequency. Using the experimental parameters, we find that $\Omega_{\text{u}}^2/4\Delta_e\sim$ 10 MHz and $\Omega'\sim$ 10 kHz while $|\Omega_{\text{l}}^2/4\Delta_e|$ is in the sub-kHz range, which can be neglected. 

We proceed by expanding the Rydberg states in terms of Floquet states $|R_j, k\rangle$ ($k=0,\pm 1,\dots$), where $k$ denotes quadrupolar excitations of the RF field. To explain the main peaks in Fig.~\ref{ryd_ex_24D}, we only need to take into account transitions $|R_1,0\rangle \to |R_2, \pm 1\rangle$, i.e. by absorbing or emitting one quadrupolar RF excitation. Eq.~(\ref{eq:Hdsim1}) becomes
\begin{eqnarray}
H_{\text{D}}&\approx&  \hbar\Delta_1 \sum_j|R_j\rangle\langle R_j|  +\frac{\hbar}{2}\left[\Omega'|R_1\rangle\langle g|+\text{H.c.}\right]  +H_{\text{F}} \nonumber, \\ 
\label{eq:Hdsim2}
\end{eqnarray}
where 
\begin{eqnarray}
H_{\text{F}}&=&- \frac{\hbar\Omega_{\text{u}}^2}{4\Delta_e}|R_1\rangle\langle R_1| +(E^{(B)}_2-E^{(B)}_1)|R_2\rangle\langle R_2| \nonumber \\
&+&\sum_k k\hbar\Omega |R_2,k\Omega\rangle\langle R_2,k\Omega|\nonumber \\
&+&\frac{\hbar C}{2}[|R_2,\Omega\rangle\langle R_1| + |R_2,-\Omega\rangle\langle R_1| + \text{H.c.}].
\end{eqnarray}

Hamiltonian $H_{\text{F}}$ is the key result. The eigenenergy of $H_{\text{F}}$ determines the peaks shown in Fig.~\ref{ryd_ex_24D}a. One example for the $\sigma^-/\sigma^+$ transition is given in Section~III. 

\begin{acknowledgments}
We thank Michael Niedermayer for electroplating the trap electrodes, and Ana Predojevi\'{c} for feedback on the manuscript. The experiment started at the University of Innsbruck and moved in 2015 to Stockholm University. Some of the experimental results were first observed in Innsbruck and have been repeated and confirmed at Stockholm University.
The research leading to these results has received funding from the European Research Council under the European Union's Seventh Framework Programme / ERC Grant Agreement No. 279508 (QuaSIRIO). 
IL acknowledges funding from the European Research Council under the European Union's Seventh Framework Programme (FP/2007-2013) / ERC Grant Agreement No. 335266 (ESCQUMA) and the H2020-FETPROACT-2014 Grant No.640378 (RYSQ). WL acknowledges access to the University of Nottingham HPC Facility.
\end{acknowledgments}

\bibliography{TrappedRydbergIon}

\end{document}